\documentclass[a4paper,fleqn,usenatbib]{mnras}

\usepackage{newtxtext,newtxmath}
\usepackage{amssymb,amsmath,latexsym,mathrsfs}
\usepackage{graphicx,subfigure}
\usepackage{epsfig}
\usepackage{hyperref}
\usepackage{varioref,xr-hyper}
\usepackage{color}
\usepackage{multirow}
\usepackage{tikz}
\usepackage{array}
\usepackage{wasysym}
\usepackage{color}
\usepackage{float}
\usepackage[utf8]{inputenc}
\usepackage[T1]{fontenc}
\usepackage{ae,aecompl}

\title[Scattering between dark energy and baryons]{Do we have any hope of detecting scattering between dark energy and baryons through cosmology?}

\author[S. Vagnozzi et al.]{
Sunny Vagnozzi,$^{1}$\thanks{E-mail: \href{mailto:sunny.vagnozzi@ast.cam.ac.uk}{sunny.vagnozzi@ast.cam.ac.uk} (SV)}\thanks{Newton-Kavli Fellow}
Luca Visinelli,$^{2}$\thanks{E-mail: \href{mailto:l.visinelli@uva.nl}{l.visinelli@uva.nl} (LV)}
Olga Mena$^{3}$\thanks{E-mail: \href{mailto:omena@ific.uv.es}{omena@ific.uv.es} (OM)}
and David F. Mota$^{4}$\thanks{E-mail: \href{mailto:d.f.mota@astro.uio.no}{d.f.mota@astro.uio.no} (DFM)}
\\
$^{1}$Kavli Institute for Cosmology, University of Cambridge, Madingley Road, Cambridge CB3 0HA, United Kingdom\\
$^{2}$Gravitation Astroparticle Physics Amsterdam (GRAPPA), University of Amsterdam, Science Park 904, 1098 XH Amsterdam, The Netherlands\\
$^{3}$Instituto de F\'{i}sica Corpuscular (IFIC), University of Valencia-CSIC, E-46980 Valencia, Spain \\
$^{4}$Institute of Theoretical Astrophysics, University of Oslo, P.O. Box 1029 Blindern, N-0315 Oslo, Norway
}

\date{Accepted XXX. Received YYY; in original form ZZZ}

\pubyear{2020}

\begin{document}
\label{firstpage}
\pagerange{\pageref{firstpage}--\pageref{lastpage}}
\maketitle

\begin{abstract}
We consider the possibility that dark energy and baryons might scatter off each other. The type of interaction we consider leads to a pure momentum exchange, and does not affect the background evolution of the expansion history. We parametrize this interaction in an effective way at the level of Boltzmann equations. We compute the effect of dark energy-baryon scattering on cosmological observables, focusing on the Cosmic Microwave Background (CMB) temperature anisotropy power spectrum and the matter power spectrum. Surprisingly, we find that even huge dark energy-baryon cross-sections $\sigma_{xb} \sim {\cal O}({\rm b})$, which are generically excluded by non-cosmological probes such as collider searches or precision gravity tests, only leave an insignificant imprint on the observables considered. In the case of the CMB temperature power spectrum, the only imprint consists in a sub-percent enhancement or depletion of power (depending whether or not the dark energy equation of state lies above or below $-1$) at very low multipoles, which is thus swamped by cosmic variance. These effects are explained in terms of differences in how gravitational potentials decay in the presence of a dark energy-baryon scattering, which ultimately lead to an increase or decrease in the late-time integrated Sachs-Wolfe power. Even smaller related effects are imprinted on the matter power spectrum. The imprints on the CMB are not expected to be degenerate with the effects due to altering the dark energy sound speed. We conclude that, while strongly appealing, the prospects for a direct detection of dark energy through cosmology do not seem feasible when considering realistic dark energy-baryon cross-sections. As a caveat, our results hold to linear order in perturbation theory.
\end{abstract}


\section{Introduction}
\label{sec:intro}

The Universe is dark and full of terrifying unknowns. Various independent astrophysical and cosmological observations~\citep[][]{Zwicky:1933gu,Rubin:1970zza,Riess:1998cb,Perlmutter:1998np,2018arXiv180706209P} indicate that most of the energy content of the Universe resides in dark matter and dark energy, whose origin and composition remain unknown. A clustering dark matter (DM) component is required to explain the inferred rotation curves of galaxies and the formation of the observed large-scale structure (LSS) of the Universe, while a smooth dark energy (DE) component is needed to explain the inferred late-time acceleration of the Universe.  For comprehensive reviews on DM and DE, see for instance~\citep{Bergstrom:2000pn,Sahni:2004ai,Bertone:2004pz,Frieman:2008sn,Bamba:2012cp}.

The leading explanation for DM posits the existence of additional particles and/or forces weakly coupled to the Standard Model~\citep[e.g.][]{Jungman:1995df, Cirelli:2005uq, ArkaniHamed:2008qn, Foot:2014uba, Hui:2016ltb}, while in principle it is possible to attribute the effects of DM to modifications of gravity~\citep[e.g.][]{Milgrom:1983ca,Chamseddine:2013kea,Rinaldi:2016oqp,Verlinde:2016toy,Vagnozzi:2017ilo}. The leading paradigm for DE consists of a cosmological constant $\Lambda$ (representing the zero-point vacuum energy density of quantum field theory), whose theoretical value is however in striking disagreement with observational inferences, an issue commonly referred to as the cosmological constant problem~\citep{Weinberg:1988cp}. At present, it is unclear whether DE is the manifestation of a new (possibly light) field~\citep[e.g.][]{Wetterich:1987fm,Ratra:1987rm,Caldwell:1997ii,Linder:2007wa,Tsujikawa:2013fta,Yang:2018xah}, a breakdown of General Relativity~\citep[e.g.][]{Li:2004rb,Nojiri:2006ri,Hu:2007nk,Myrzakulov:2015qaa,Sebastiani:2016ras}, or something else altogethether~\citep[for instance][]{Rasanen:2003fy,Rinaldi:2014yta,Nunes:2016aup}.~\footnote{Although DM and DE are in principle independent components, each evolving following a separate continuity equation, an interesting possibility considered in the literature is that the two components might interact with each other, as occurring in so-called interacting dark energy models. For examples of such models, see e.g.~\citep{Wetterich:1994bg,Amendola:1999er,Farrar:2003uw,dfm1,dfm2,Gavela:2009cy,Pan:2012ki,Tamanini:2015iia,Yang:2017zjs,Yang:2017ccc}, and see instead~\citep{Wang:2016lxa} for a review.}

Is it possible to investigate the physics of the dark sector in the laboratory? A huge experimental program has been dedicated to detecting and characterizing the nature of DM. The leading search strategies can essentially be divided into three categories: \textit{collider production} of DM, \textit{indirect detection} of the products of DM annihilation or decay, and \textit{direct detection} of collisions between DM and target nuclei: see e.g.~\citep{Kahlhoefer:2017dnp,Gaskins:2016cha,Undagoitia:2015gya} for reviews on these three experimental strategies. The goal of direct detection experiments is to search for the recoil energy deposited in collisions between galactic halo DM particles and target nuclei in a detector~\citep{Goodman:1984dc}. In addition, one expects the signal to be modulated on the scales of both a sidereal year~\citep{Drukier:1986tm,Freese:2012xd} and a sidereal day~\citep{Collar:1992qc,Kouvaris:2014lpa,Foot:2014osa} due to the revolution of the Earth around the Sun and around its axis. Examples of leading direct detection experiments include (but are not limited to) CoGeNT~\citep{Aalseth:2012if}, CRESST~\citep{Petricca:2017zdp}, DAMA-LIBRA~\citep{Bernabei:2008yh}, LUX~\citep{Akerib:2013tjd}, and XENON1T~\citep{Aprile:2018dbl}. Alternative direct detection strategies involving among others ancient minerals~\citep{2018arXiv180605991B,Drukier:2018pdy,Edwards:2019puy}, superfluid Helium~\citep{Schutz:2016tid}, and DNA~\citep{Drukier:2014rea}, are also being proposed.

At present, no parallel search for the (direct) detection of dark energy is being carried out. Most of the motivation resides in the fact that, while for the DM there exist several viable candidates such as WIMPs~\citep{Roszkowski:2017nbc}, axions~\citep{Abbott:1982af,Dine:1982ah,Visinelli:2009zm}, primordial black holes~\citep{Hawking:1971ei, Carr:2016drx}, or sterile neutrinos~\citep{Dodelson:1993je, Boyarsky:2018tvu}, the situation with DE is much less clear, since we do not even know whether the latter is a manifestation of a theory of gravity beyond General Relativity, possibly in connection to string theory~\citep{Banerjee:2018qey}, or the existence of new fields. In the latter case, it is not even clear what the associated mass scale should be.~\footnote{For instance, if DE is in the form of a very light axion~\citep[e.g. as in][]{Arvanitaki:2009fg, Hlozek:2014lca, Visinelli:2018utg}, its mass $m_a$ has to be of order the Hubble rate today, $m_a \sim H_0 \sim 10^{-33}\,$eV, in order for Hubble friction to efficiently freeze the motion of the particle and achieve an effective cosmological constant-like equation of state.} Collider searches for DE have been studied in detail in a very limited number of works~\citep{Brax:2009aw, Brax:2009ey, Brax:2014vva, Brax:2015hma, Brax:2016did}, including those recently carried out by the ATLAS collaboration~\citep{Aaboud:2019yqu}.

Most of the searches for DE properties have been conducted on the cosmological side. These have focused on the DE equation of state (EoS) $w_x$ and its time evolution, or on models of modified gravity that can account for DE, by studying imprints on the background evolution and on the late-time growth of structure~\citep[see e.g.][]{Ishak:2005zs,Mena:2005ta,DeFelice:2007ez,Giannantonio:2009gi,Lombriser:2010mp,Martinelli:2011wi,Hu:2016zrh,Nunes:2016plz,Nunes:2016drj,Renk:2017rzu,Peirone:2017vcq,Vagnozzi:2018jhn,Casalino:2018mna,Du:2018tia,Yang:2019vni}, and finally on the propagation of astrophysical gravitational waves~\citep[see e.g.][]{Creminelli:2017sry,Sakstein:2017xjx,Ezquiaga:2017ekz,Boran:2017rdn,Baker:2017hug,Visinelli:2017bny,Crisostomi:2017lbg,Langlois:2017dyl,Ezquiaga:2018btd,Casalino:2018wnc}. Future DE surveys such as DESI~\citep{2016arXiv161100036D}, Euclid~\citep{Amendola:2012ys}, WFIRST~\citep{2015arXiv150303757S}, and LSST~\citep{2012arXiv1211.0310L} will use combinations of galaxy clustering, weak lensing, redshift-space distortions, and cross-correlations between all these probes. These will substantially improve our understanding of DE and may rule out the cosmological constant $\Lambda$. However, much remains to be understood about DE. One relevant example is related to the $H_0$ tension, the mismatch between high- and low-redshift determinations of the Hubble constant $H_0$~\citep{2018arXiv180706209P,Riess:2019cxk}. It has been argued that such tensions might be eased by introducing non-minimal physics in the dark energy sector~\citep{2019arXiv191009853D}, in the form of phantom DE~\citep{DiValentino:2016hlg,Bernal:2016gxb,2019arXiv190707569V}, DM-DE interactions~\citep{dfm3,dfm4,DiValentino:2017iww,Yang:2018euj,Yang:2018uae,
Martinelli:2019dau,Kumar:2019wfs,2019arXiv190804281D},~\footnote{Note, however, that a possible drawback in this case is that the reason why the $H_0$ tension is alleviated is often due to increased degeneracies between cosmological parameters, which broaden all the constraints including the ones on $H_0$. Therefore, introducing additional datasets which can break these degeneracies (usually low-redshift datasets) typically leads to the $H_0$ tension reappearing~\citep[see e.g.][]{Martinelli:2019dau}.} an early DE component~\citep{Karwal:2016vyq,Mortsell:2018mfj,Poulin:2018cxd,2019arXiv190401016A}, vacuum metamorphosis~\citep{DiValentino:2017rcr}, or running vacuum~\citep{Sola:2017znb,Gomez-Valent:2018nib,Rezaei:2019xwo,Sola:2019jek}, whereas other possibilities such as a phase transition in the DE~\citep{DiValentino:2019exe} or a negative cosmological constant~\citep{2018arXiv180806623D,Visinelli:2019qqu} do not appear to alleviate the problem.

The current ``asymmetry'' between experimental searches and cosmological surveys for investigating DM and DE has motivated the present study, aiming at answering the following questions: what if DE scattered off baryons? What signature would such an interaction leave on cosmological observables, as the CMB or the distribution of the LSS? Is cosmological \textit{direct detection of dark energy}, analogous to direct detection of dark matter, even remotely possible? Admittedly, adding interactions between DE and baryons is a risky procedure, since such a new interaction might lead to long-range forces~\citep{Brax:2019iut} and variations in the fundamental constants~\citep{Uzan:2002vq}, which are severely restricted by observations and potentially dangerous~\citep{Barrow:1998df,Barrow:2002ed,Mota:2003tm,Uzan:2010pm,2017RPPh...80l6902M}. While this threat certainly holds for specific DE models such as quintessence~\citep[as shown in][]{Carroll:1998zi}, it does not have to be a danger in general. On a more fundamental level, (scalar) fields which might play the role of DE are ubiquitous in extensions of the Standard Model (SM). In this case, it is inevitable that they will couple to baryons to some extent, either through a direct tree-level Lagrangian coupling or at the loop level (unless such a coupling is forbidden by a fundamental symmetry), or indirectly through the Ricci scalar~\citep[see e.g.][]{Wetterich:1987fm,Damour:1990tw,Biswas:2005wy}. Therefore, from a field theory perspective it is quite hard to imagine how DE and baryons can be completely decoupled.

Earlier analyses considered the possibility of scattering between DM and DE and showed that in principle rather large DM-DE scattering cross-sections are allowed by cosmological data~\citep{Simpson:2010vh}. In that work, it was also conjectured that scattering between DE and baryons is allowed with very large cross-section, albeit such a conjecture was not further justified. Another related earlier work is that of~\citep{Calabrese:2013lga}, which studied the cosmology of DE interacting with the electromagnetic sector of the SM of elementary particles. Motivated by such questions, one of our major goals in this work is to check whether such a speculation over a large DE-baryon interaction is indeed correct. We will therefore allow for DE and baryons to scatter. We introduce an effective scattering term between the two components at the level of Boltzmann equations. Certainly, such a scattering progress can be expected to lead to changes in cosmological observables, and our goal here is to understand what these changes are, and whether such changes might in principle be visible in current or future surveys, opening up a new window onto the physics of DE. In other words, our aim is to undertake the possibility of a cosmological direct detection of dark energy.

The rest of this paper is then organized as follows. In Sec.~\ref{sec:interactions} we describe how the standard Boltzmann equations are modified in the presence of a dark energy-baryon scattering process. We also provide a rough estimate for how large the scattering cross-section can be given current non-cosmological probes, such as collider searches. In Sec.~\ref{sec:results} we discuss how standard cosmological probes such as the CMB and matter power spectra are modified in the presence of a DE-baryon scattering. For pedagogical purposes, in order to boost the effect of the dark energy-baryon scattering, we will focus on extremely large cross-sections, of order barn, where recall the barn is defined as $1\,{\rm b} \equiv 10^{-24}\,{\rm cm}^2$. We will show that even such large cross-sections lead to tiny modifications to the cosmological observables. Sec.~\ref{sec:physicalexplanation} presents the physics behind the imprints of dark energy-baryon scattering on cosmological observables. In the case of the CMB, these are directly related to changes in the late-time integrated Sachs-Wolfe effect. Finally, we conclude in Sec.~\ref{sec:conclusions} summarizing our results and discussing future prospects.

\section{Dark energy-baryon scattering}
\label{sec:interactions}

We begin by discussing how the standard Boltzmann equations are modified in the presence of a DE-baryon scattering process. We then provide rough estimates of how large the scattering cross-section is allowed to be given current non-cosmological probes, focusing on collider searches for dark energy as well as on precision tests of gravity.

\subsection{Boltzmann equations}
\label{subsec:boltzmann}

We work in the synchronous gauge~\citep{Lifshitz:1945du}, which is the gauge adopted by the Boltzmann solver \texttt{CAMB}~\citep{Lewis:1999bs}. In this gauge, the perturbed Friedmann-Lema\^{i}tre-Robertson-Walker (FLRW) line element is given by:
\begin{eqnarray}
ds^2 = a^2(\eta) \left [ -d\eta^2 + (\delta_{ij}+h_{ij})dx^idx^j \right ]\,,
\end{eqnarray}
with $\eta$ denoting conformal time. Within this gauge, our goal is to track the evolution of the Fourier-space baryon density contrast and the velocity divergence $\delta_b$ and $\theta_b$, and the DE density contrast and the velocity divergence $\delta_x$ and $\theta_x$, in the presence of a DE-baryon scattering process, characterized by a cross-section $\sigma_{xb}$ quantifying the strength of DE-baryon scattering.

A few comments are in order at this point. We will be considering a purely elastic scattering process, \textit{i.e.} a process in which there is no energy transfer coming along with momentum transfer. Consequently, these models are quite different from the interacting DM-DE models we described in Sec.~\ref{sec:intro}, where energy transfer occurs between DM and DE, and the background evolution is modified by such a process. In contrast, in our scenario the background evolution remains unaltered, whereas it is only the evolution of perturbations which is affected. As a further clarification, when we say that the background evolution remains unaltered, what we mean is that it is unaltered with respect to the background evolution in the same cosmology without DE-baryon scattering. The latter may or may not be a $\Lambda$CDM cosmology. In fact, in this work we will only consider non-$\Lambda$CDM cosmologies where the DE EoS is $w_x \neq -1$ (so-called $w$CDM cosmologies). The reason, as we shall see later when we write down the Boltzmann equations in Eqs.~(\ref{eq:deltab}-\ref{eq:thetade}), is that only when $w_x \neq -1$ can DE-baryon scattering modify the Boltzmann equations. To put it differently, baryons cannot scatter off a cosmological constant. The background evolution in a $w$CDM cosmology with DE-baryon scattering is the same as that in the original $w$CDM cosmology without DE-baryon scattering.

Such a scenario may be argued to be unnatural. However, given the low DE density and the non-relativistic baryon velocities, such an interaction could be regarded as natural. In general, elastic scattering can be a valid approximation as long as the mass/energy scale of one particle (in this case that associated to the DE component, for which we expect $E \ll {\cal O}({\rm eV})$) is much smaller than the mass/energy scale of the other (in this case baryons, for which $E \gg {\cal O}({\rm eV})$), as is the case for instance with Thomson scattering. Similar examples of elastic scattering with dark sector components have been considered in the literature, for instance in the context of DM-photon scattering~\citep{Wilkinson:2013kia, Stadler:2018jin,Kumar:2018yhh}, DM-neutrino scattering~\citep{Mangano:2006mp,Serra:2009uu,Wilkinson:2014ksa,Escudero:2015yka, Stadler:2019dii}, DM-DE scattering~\citep{Simpson:2010vh, Xu:2011nr, Baldi:2014ica,Skordis:2015yra, Baldi:2016zom, Kumar:2017bpv, Asghari:2019qld}, DM-baryon scattering~\citep{Gluscevic:2017ywp,Boddy:2018kfv,Xu:2018efh,Fialkov:2018xre,Boddy:2018wzy}, and DM-DM scattering~\citep{Cyr-Racine:2015ihg,Vogelsberger:2015gpr,Archidiacono:2017slj,Buen-Abad:2017gxg,Archidiacono:2019wdp}. We follow a purely phenomenological approach here: as argued in earlier works~\citep{Simpson:2010vh, Xu:2011nr}, this choice is justified  since the macrophysics involved in shaping the cosmological observables we are interested in is expected to be, to a more than reasonable approximation, independent of the microphysics involved in the scattering process.

With the above discussion in mind, we now modify the standard Boltzmann equations to account for an elastic DE-baryon scattering process quantified by a cross-section $\sigma_{xb}$, in the spirit of what was done earlier in~\citep{Simpson:2010vh} and later in~\citep{Xu:2011nr, Baldi:2014ica,Skordis:2015yra, Baldi:2016zom, Kumar:2017bpv, Asghari:2019qld} for elastic DE-DM scattering. In the synchronous gauge, these equations read:
\begin{eqnarray}
\dot{\delta}_b &=& -\theta_b-\frac{\dot{h}}{2}\,,\label{eq:deltab} \\
\dot{\theta}_b &=& -{\cal H}\theta_b + c_s^2k^2\delta_b + \frac{4\rho_{\gamma}}{3\rho_b}an_e\sigma_T(\theta_{\gamma}-\theta_b) \nonumber \\
&&+(1+w_x)\frac{\rho_x}{\rho_b}an_e\sigma_{xb}(\theta_x-\theta_b) \,,\label{eq:thetab} \\
\dot{\delta}_x &=& -(1+w_x) \left ( \theta_x+\frac{\dot{h}}{2} \right )-3{\cal H}(c_{s,x}^2-w_x)\delta_x \nonumber \\
&& -9{\cal H}^2(c_{s,x}^2-w_x)(1+w_x)\frac{\theta_x}{k^2} \,, \label{eq:deltade} \\
\dot{\theta}_x &=& -{\cal H}(1-3c_{s,x}^2)\theta_x+\frac{c_{s,x}^2k^2}{1+w_x}\delta_x+an_e\sigma_{xb}(\theta_b-\theta_x)\,.\label{eq:thetade}
\end{eqnarray}
Here, $h$ is the usual synchronous gauge metric perturbation~\citep[see e.g.][]{Ma:1995ey} and $\sigma_T \approx 6.7 \times 10^{-25}\,{\rm cm}^2 = 0.67\,{\rm b}$ is the Thomson scattering cross-section between baryons and photons (where $1\,{\rm b} = 10^{-24}\,{\rm cm}^2$ defines the barn measurement unit). Moreover, $\rho_b$, $\rho_{\gamma}$, and $\rho_x$ are the baryon, photon, and DE energy densities respectively, $w_x$ is the DE equation of state, $c_s^2$ is the baryon sound speed squared, and $c_{s,x}^2$ is the DE sound speed squared. Finally, $a$ is the scale factor and $n_e$ is the number density of electrons.

With little loss of generality, we shall fix from now on $c_{s,x}^2=1$. In writing Eqs.~(\ref{eq:deltade}) and~(\ref{eq:thetade}) we have also set the time variation of the DE equation of state to zero, so that the DE adiabatic sound speed squared reads $c_a^2=w_x$~\citep{Hu:1998kj}. Note that the equations for the density contrasts are unaltered, given that a momentum transfer process is only expected to modify the equations for the velocity divergences. Of course, any modification to the baryon and DE velocities due to the scattering will in turn modify the density contrasts by backfeeding into Eqs.~(\ref{eq:deltab}) and~(\ref{eq:deltade}). When introducing the DE-baryon scattering term in Eq.~(\ref{eq:thetab}), the pre-factor $(1+w_x)\rho_x/\rho_b$ has been introduced in order to conserve total momentum during the DE-baryon scattering, as expected during an elastic scattering process. This is completely analogous to the $4\rho_{\gamma}/3\rho_b$ pre-factor appearing in Eq.~(\ref{eq:deltab}) when describing Thomson scattering between baryons and photons~\citep[see e.g.][]{Ma:1995ey}. The new term in Eqs.~(\ref{eq:thetab}) and~(\ref{eq:thetade}) is effectively describing a drag term for the DE velocity, with $n_e\sigma_{xb}(\theta_b-\theta_x)$ representing the fraction of DE quanta which are subject to scattering off baryons per unit time. The above equations also clarify our earlier statement that, when $w_x=-1$ (\textit{i.e.} when DE is in the form of a cosmological constant), DE and baryons cannot scatter. In fact, the DE-baryon scattering term in Eq.~(\ref{eq:thetab}) shuts off when $w_x=-1$, due to the $(1+w_x)$ pre-factor. Moreover, a cosmological constant is smooth and does not feature perturbations, thus Eq.~(\ref{eq:thetade}) is not tracked when $w_x=-1$.

Our aim is now to implement the modified Boltzmann equations described above in a Boltzmann solver, such as \texttt{CAMB}~\citep{Lewis:1999bs}. Following standard notation~\citep[e.g.][]{Ma:1995ey}, we define the photon-to-baryon density ratio $R \equiv 4\rho_{\gamma}/3\rho_b$ and the Thomson scattering opacity $\tau_c \equiv (an_e\sigma_T)^{-1}$. Analogously, we define the DE-to-baryon density ratio $R_x \equiv (1+w_x)\rho_x/\rho_b$ and the DE-baryon scattering opacity $\tau_x \equiv (an_e\sigma_{xb})^{-1}$. It is numerically convenient to work with the dimensionless quantity $\alpha_{xb}$ given by the ratio of the DE-baryon interaction cross-section to the Thomson cross-section, $\alpha_{xb} \equiv \sigma_{xb}/\sigma_T = \tau_c/\tau_x$. We refer to $\alpha_{xb}$ as the ``Thomson ratio''. Having defined these quantities, and setting $c_{s,x}^2=1$ as we discussed earlier, we can now rewrite Eqs.~(\ref{eq:thetab}) and~(\ref{eq:thetade}) as:
\begin{eqnarray}
\dot{\theta}_b \!&=&\! -{\cal H}\theta_b + c_s^2k^2\delta_b + R\tau_c^{-1}(\theta_{\gamma}-\theta_b)+R_x\tau_c^{-1}\alpha_{xb}(\theta_x-\theta_b) \,,\nonumber \\ \label{eq:thetabnew} \\
\dot{\theta}_x \!&=&\! 2{\cal H}\theta_x+\frac{k^2}{1+w_x}\delta_x+\tau_c^{-1}\alpha_{xb}(\theta_b-\theta_x) \,. \label{eq:thetadenew}
\end{eqnarray}
The most immediate question at this point is: what values of $\alpha_{xb}$ are allowed by non-cosmological probes such as collider searches or precision gravity tests? We will address this question in Sec.~\ref{subsec:howlarge}, before moving on to the cosmological constraints presented in Sec.~\ref{sec:results}.

\subsection{How large can the dark energy-baryon cross-section be?}
\label{subsec:howlarge}

In this section we estimate the size, allowed by non-cosmological probes, of the DE-baryon cross-section or, equivalently, of the Thomson ratio $\alpha_{xb}$. While a model-independent bound is very hard to derive, we shall argue that on quite general grounds one can generically expect $\alpha_{xb} \ll 1$ to hold. We remind the reader that $\alpha_{xb} \simeq {\cal O}(1)$ would correspond to a large nuclear-scale cross-section, of order barn and comparable to the Thomson cross-section. If such a large DE-baryon interaction exists, it will be extremely hard to conceive how it might have escaped detection. For instance, it is hard to imagine how such a process would not have been seen at colliders, or even in experiments devoted to the direct detection of DM. In the latter case, even though the local DE density is much lower than the local DM density, this would be completely offset by the much larger cross-section. The only possibility would be if such an interaction were screened, for instance through the Vainshtein mechanism~\citep{Vainshtein:1972sx} which invokes non-linearities in the vicinity of matter sources. However, again it would be challenging to screen such a large interaction, especially on cosmological scales, where gravitational potentials are typically much smaller than those available locally.

Let us be more concrete and study possible collider limits on the Thomson ratio $\alpha_{xb}$. Focusing on the rather general case where DE is described by a scalar field $\phi$ with a mass comparably smaller than collider energy scales, one might consider an effective field theory (EFT) description of all possible interactions between $\phi$ and the SM particles. Such an approach was pioneered in a number of works, including the very comprehensive analysis of~\citep{Brax:2016did}, which takes into account a set of effective operators containing several well-known DE models as a subset, such as chameleon DE, symmetron DE, quintessence, and galileons. It is worth warning the reader that an EFT approach is strictly speaking valid only if the energy scales probed in the relevant experiment are far from the scale where new physics comes into play, or in other words if resonances and thresholds are unresolved. Should this not be the case, one needs to revert to specific UV completions. In our case, we expect new gravitational physics (which would complicate the analysis significantly due to the non-linearity and non-renormalizability of gravity) to come into play well above the TeV scale probed by the LHC, making an EFT approach somewhat justified.~\footnote{We wish to clarify to the reader that the EFT approach we are discussing is distinct from, albeit related to, the EFT of DE description developed in~\citep{Gubitosi:2012hu,Bloomfield:2012ff,Piazza:2013coa}. The latter is used to describe all single-field DE and modified gravity models in terms of the most general action written in unitary gauge and considering operators compatible with residual symmetries of unbroken spatial diffeomorphism, along the spirit of the EFT of inflation~\citep{Cheung:2007st}, and is implemented in Boltzmann solvers such as \texttt{EFTCAMB}~\citep{Hu:2013twa,Raveri:2014cka}. See e.g.~\citep{2019arXiv190703150F} for a review on the EFT of DE.} In the following, we shall follow this approach and consider for concreteness two specific EFT operators.

Following~\citep{Brax:2016did,Aaboud:2019yqu}, we consider shift-symmetric EFT operators (\textit{i.e.} EFT operators invariant under a transformation $\phi \to \phi+{\rm const}$), which thus couple to the SM through derivative interactions. Shift-symmetry breaking operators are instead tightly constrained by precision gravity tests and thus are not expected to leave signatures in colliders, nor to be cosmologically relevant~\citep{Joyce:2014kja}. There are nine shift-symmetric operators, but for conciseness we consider only the two leading ones, \textit{i.e.} the two least suppressed. These two dimension-8 operators, ${\cal L}_1$ and ${\cal L}_2$, are usually referred to as kinetically dependent conformal and disformal operators respectively, and are given by:
\begin{eqnarray}
{\cal L}_1 &=& \frac{\partial_{\mu}\phi\partial^{\mu}\phi}{M_1^4}T_{\nu}^{\nu} \,, \label{eq:operatorell1} \\
{\cal L}_2 &=& \frac{\partial_{\mu}\phi\partial_{\nu}\phi}{M_2^4}T^{\mu \nu} \,, \label{eq:operatorell2}
\end{eqnarray}
where $M_1$ and $M_2$ are two suppression scales and $T_{\mu \nu}$ is the energy-momentum tensor of the SM fields. The first operator couples to the trace of the energy-momentum tensor and hence to the conformal anomaly. Of interest to us here will be fermionic fields $\psi_i$ of mass $m_i$, for which $T^{\nu}_{\nu}=m_i\Bar{\psi}_i\psi_i$. The most sensitive production channel for the first operator thus involves DE production in association with $t\Bar{t}$ (as the top quark is heaviest fermion of the SM), whereas the most sensitive production channel for the second operator involves jets and missing transverse energy, given that the coupling to the energy-momentum tensor of the SM fields implies that the production cross-section will be proportional to their momenta.

The analyses of~\citep{Aaboud:2019yqu} set limits of about $M_1 \gtrsim 200\,{\rm GeV}$ and $M_2 \gtrsim 1.2\,{\rm TeV}$, whereas the earlier results of~\citep{Brax:2016did} with less data had set weaker but comparable limits. It is then instructive to compute the typical cross-sections associated to the operators in Eqs.~(\ref{eq:operatorell1}) and~(\ref{eq:operatorell2}), which we refer to as $\sigma_1$ and $\sigma_2$ respectively. Up to factors of order unity or at most ${\cal O}(10)$, irrelevant for the subsequent discussion, we find:
\begin{eqnarray}
\sigma_1 \sim \frac{p_{\phi}^4m_i^2}{M_1^8}\,, \label{eq:sigma1} \\
\sigma_2 \sim \frac{p_{\phi}^4p_i^2}{M_2^8}\,, \label{eq:sigma2}
\end{eqnarray}
where $p_{\phi}$ denotes the DE momentum, $m_i$ denotes the mass of the SM particle produced in association with the DE, and similarly $p_i$ denotes the momentum of the SM particle or jet produced in association with the DE. Inserting numbers into Eqs.~(\ref{eq:sigma1}) and~(\ref{eq:sigma2}) and using the upper limits on $M_1$ and $M_2$ derived in~\citep{Aaboud:2019yqu} we notice that, even in the most optimistic scenario, the relevant cross-section is going to be well below the barn scale. In other words, this implies that $\alpha_{xb} \ll 1$. More concretely in~\citep{Brax:2015hma,Aaboud:2019yqu} it is shown that ATLAS and CMS exclude production cross-sections in relation to the two operators in question of order pb-fb. This means that we can expect $\alpha_{xb} \ll 10^{-12}$ from the non-observation of DE in colliders.

Two comments are in order at this point. The considerations we have made are strictly speaking only valid for the two EFT operators of Eqs.~(\ref{eq:operatorell1}) and~(\ref{eq:operatorell2}). We expect nonetheless that they should extend quite generally to many realistic DE models, given the non-observation of DE at colliders, the fact that the effective operators considered contain several well-known DE models as a subset, and the two operators we considered are the two least-suppressed, as considering more suppressed operators would only strengthen our conclusions. The second comment is that the rough upper limits on the Thomson ratio $\alpha_{xb}$ we have derived strictly speaking would only hold on $\approx {\rm TeV}$ scales. They can be safely extrapolated to the energy scales relevant for cosmology only insofar as we do not expect the DE-baryon coupling to run significantly with energy. It is not possible to quantify whether one should expect a significant running in the absence of a UV complete model, but again we generically expect that $\alpha_{xb} \ll 1$ should hold.

So far we have discussed collider searches, but what about other non-cosmological tests? In~\citep{Mota:2006fz} the authors have argued that a combination of terrestrial tests such as the Eot-Wash experiment~\citep{Hoyle:2000cv,Kapner:2006si,Wagner:2012ui}, measurements of the Casimir force, constraints from the physics of compact objects (such as white dwarfs), weak equivalence principle violation constraints, and precision tests of gravity within the solar system, generically lead to the expectation that the dimensionless coupling between DE and baryons should be weaker than $10^{-5}$. This leads again to the expectation that $\alpha_{xb} \ll 1$. As shown in~\citep{2013arXiv1309.5389J}, on Mpc scales one also expects the dimensionless coupling between DE and baryons to be weaker than $10^{-5}$, and hence again $\alpha_{xb} \ll 1$~\citep[see also][]{Burrage:2017qrf}.

While, as stated in the beginning of this section, a model-independent constraint is extremely hard (if not impossible) to derive, non-cosmological probes, such as our non-observation of DE in colliders or in precision tests of gravity, robustly establish that $\alpha_{xb} \ll 1$. The question we shall address is then the following one: can cosmology give us more information on the Thomson ratio $\alpha_{xb}$? As we shall illustrate we find that the answer is, very surprisingly and quite unfortunately, no. We find that $\alpha_{xb} \sim {\cal O}(1)$ or even $\alpha_{xb} \gg {\cal O}(1)$, at strong odds with non-cosmological limits, is completely consistent with current measurements of CMB anisotropies and the clustering of the large-scale structure.

\section{Results}
\label{sec:results}

Here, we discuss the impact of varying the Thomson ratio $\alpha_{xb}$ on the CMB temperature anisotropy power spectrum and the matter power spectrum. Before doing so we need to take a closer look at Eqs.~(\ref{eq:thetab}) and~(\ref{eq:thetade}). From Eq.~(\ref{eq:thetab}) we see that if we consider a cosmological constant $\Lambda$, for which $w_x=-1$, then DE-baryon scattering has no effect on the baryon velocity equation. This is a consequence of the conservation of momentum leading to the appearance of the $(1+w_x)$ factor in the DE-to-baryon energy density ratio $R_x$. Moreover, a pure cosmological constant has a smooth value over our Hubble patch, without spatial perturbations. These considerations imply that we need to move beyond $w_x=-1$ if we want to see an impact of DE-baryon scattering on cosmological observables. In particular, the effect of the DE-baryon scattering would depend on the sign of $(1+w_x)$, in other words on whether the DE EoS lies in the quintessence regime ($w_x>-1$) or in the phantom regime ($w_x<-1$).

We begin by considering the case where the DE EoS lies in the quintessence regime, \textit{i.e.} $w_x>-1$, which is somewhat theoretically favoured over the phantom regime. We find that for physically acceptable values of the Thomson ratio $\alpha_{xb} \ll 1$ as discussed in Sec.~\ref{subsec:howlarge}, adding a DE-baryon coupling leads to virtually no effect on the CMB temperature power spectrum. We could in principle try to boost the effect of the DE-baryon scattering by allowing the DE EoS $w_x$ to deviate significantly from $-1$. However, this is not a reasonable approach to take, given that late-time measurements of the expansion rate from Baryon Acoustic Oscillations (BAO) and Type-Ia Supernovae, in combination with CMB measurements, tightly constrain the DE EoS to be close to that of the cosmological constant $\Lambda$. How much of a deviation from $w_x=-1$ can be tolerated is to some extent a data-dependent statement, depending for instance on which BAO measurements one chooses to adopt, whether one chooses to include a prior on the local measurement of $H_0$, or whether one includes CMB lensing reconstruction measurements. Recall that the CMB in itself is a poor probe of $w_x$ due to the well-known geometrical degeneracy, \textit{i.e.} the fact that at linear level (without including CMB lensing) cosmological models with identical spectra of fluctuations, matter content, and angular diameter distance to last-scattering will lead to nearly indistinguishable CMB spectra~\citep[see e.g.][]{Efstathiou:1998xx}. It follows that, at least in principle, CMB measurements alone allow for rather extreme values of $w_x$, particularly in the phantom region. Indicatively, combinations of CMB and late-time measurements can approximately tolerate deviations from the cosmological constant EoS of $\vert \Delta w_x \vert \approx 0.2$, see for instance the discussions in Sec.~7.4 of~\citep{2018arXiv180706209P}.

Therefore, for purely pedagogical purposes, to boost the DE-baryon scattering signal as much as possible while not upsetting late-time measurements too much, we begin by fixing the DE EoS to $w_x=-0.8$. Still for purely pedagogical purposes we then consider values of $\alpha_{xb}$ of ${\cal O}(1)$ or slightly smaller, which we remind the reader are in principle in tension with non-cosmological measurements. Our rationale is that if we can show that, for $\alpha_{xb} \sim {\cal O}(1)$, even in the most optimistic case (by maximizing $\Delta w_x$ consistently with what is allowed by current cosmological datasets), the resulting changes in the CMB and matter power spectrum are too small to ever be observed, we can conclude that cosmological direct detection of dark energy will not be feasible.

\begin{figure}
\includegraphics[width=0.5\textwidth]{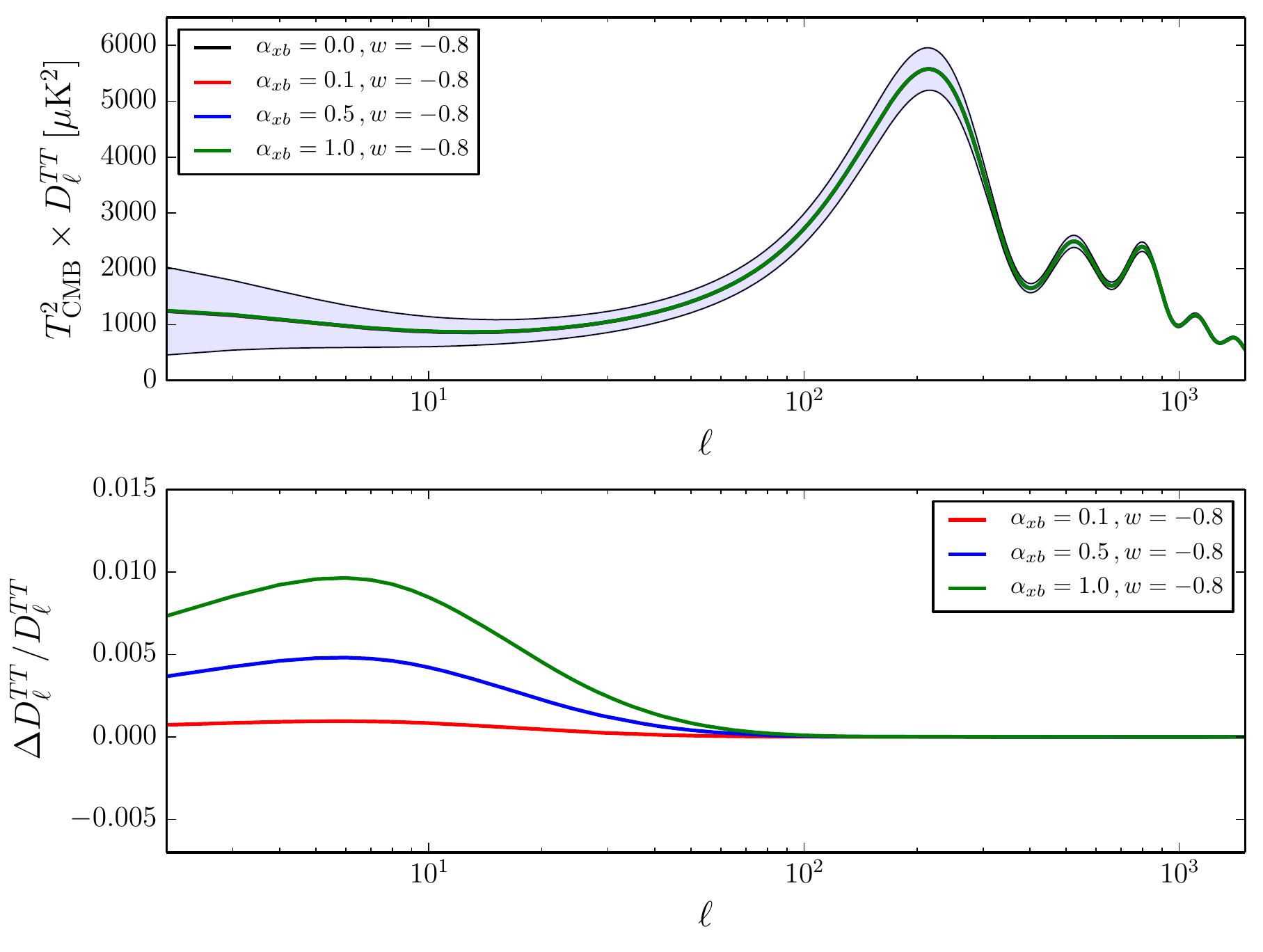}
\caption{Impact of increasing the Thomson ratio $\alpha_{xb}$, which gives the ratio between the DE-baryon scattering cross-section to the Thomson cross-section, on the CMB temperature power spectrum. \textit{Upper panel}: CMB temperature anisotropy power spectra for $\alpha_{xb} = 0$ (black curve), $0.1$ (red curve), $0.5$ (blue curve), and $1.0$ (green curve). All cosmological parameters are fixed to their best-fit values given the \textit{Planck} 2018 results~\citep{2018arXiv180706209P}, with the exception of the DE EoS which is fixed to $w_x=-0.8$, in the quintessence region. Notice that, as per standard convention in the field, the quantity plotted on the $y$ axis is $T_{\rm CMB}^2D_{\ell} \equiv T_{\rm CMB}\ell(\ell+1)C_{\ell}$, with $T_{\rm CMB} \approx 2.725\,{\rm K}$ the CMB temperature today. The light blue band indicates the uncertainty budget arising from cosmic variance, which gives $\Delta D_{\ell}/D_{\ell} \sim 1/f_{\rm sky}\sqrt{2/(2\ell+1)}$, where we have set the observed sky fraction to $f_{\rm sky}=1$. \textit{Lower panel}: since the differences between the different curves in the upper panel are too small to be appreciated by the naked eye, we plot the relative change in power with respect to the baseline model with $\alpha_{xb}=0.0$, with the same color coding as above. The increase in power at low multipoles is due to an enhanced late-time integrated Sachs-Wolfe effect, as we explain in Sec.~\ref{sec:physicalexplanation}.}
\label{fig:fig1}
\end{figure}

Our results are shown in Fig.~\ref{fig:fig1}, where we compare the resulting CMB temperature anisotropy power spectra to the baseline power spectrum obtained for the case with no DE-baryon interactions ($\alpha_{xb}=0$). As the upper panel of Fig.~\ref{fig:fig1} clearly shows, the changes induced by the DE-baryon scattering, even for the extreme case $\alpha_{xb}=1$, are not appreciable by the naked eye. The lower panel of Fig.~\ref{fig:fig1} shows the relative deviations of the resulting CMB spectra from the baseline case, showing that scattering between a quintessence-like DE component and baryons enhances the CMB power spectrum by at most ${\cal O}(1\%)$. In fact, we find that for $w_x=-0.8$ the maximum relative change in the power spectrum occurs at $\ell \approx 5$ and is well approximated by $\Delta C_{\ell}/C_{\ell}\vert_{\rm max} \approx \alpha_{xb}/100$. We have checked that this holds even for more extreme values of $\alpha_{xb} \gtrsim {\cal O}(10)$, whose results for the sake of conciseness we do not show here.

Leaving aside the fact that $\alpha_{xb} \sim {\cal O}(1)$ is unrealistically large as we argued in Sec.~\ref{subsec:howlarge}, detecting changes as small as those shown in Fig.~\ref{fig:fig1} is by all means impossible. In fact, the variations in the CMB power spectrum with respect to the baseline $\alpha_{xb}=0$ cosmology occur at extremely large angular scales (low multipoles $\ell$), where cosmic variance completely dominates and blows up the measurement error bars~\citep[see for instance][]{2018arXiv180706209P}, very strongly undermining any hope of seeing such a signal. See the light blue band in the upper panel of Fig.~\ref{fig:fig1} for the contribution of cosmic variance to the measurement uncertainty, which goes as $\Delta C_{\ell}/C_{\ell} \sim \sqrt{2/(2\ell+1)}$ for a full-sky survey. It is nonetheless interesting to consider the physics underlying the changes we find in Fig.~\ref{fig:fig1}. We expect them to be due to a change in the strength of the late-time integrated Sachs-Wolfe (LISW) effect, and will discuss these further in Sec.~\ref{sec:physicalexplanation}.

\begin{figure}
\includegraphics[width=0.5\textwidth]{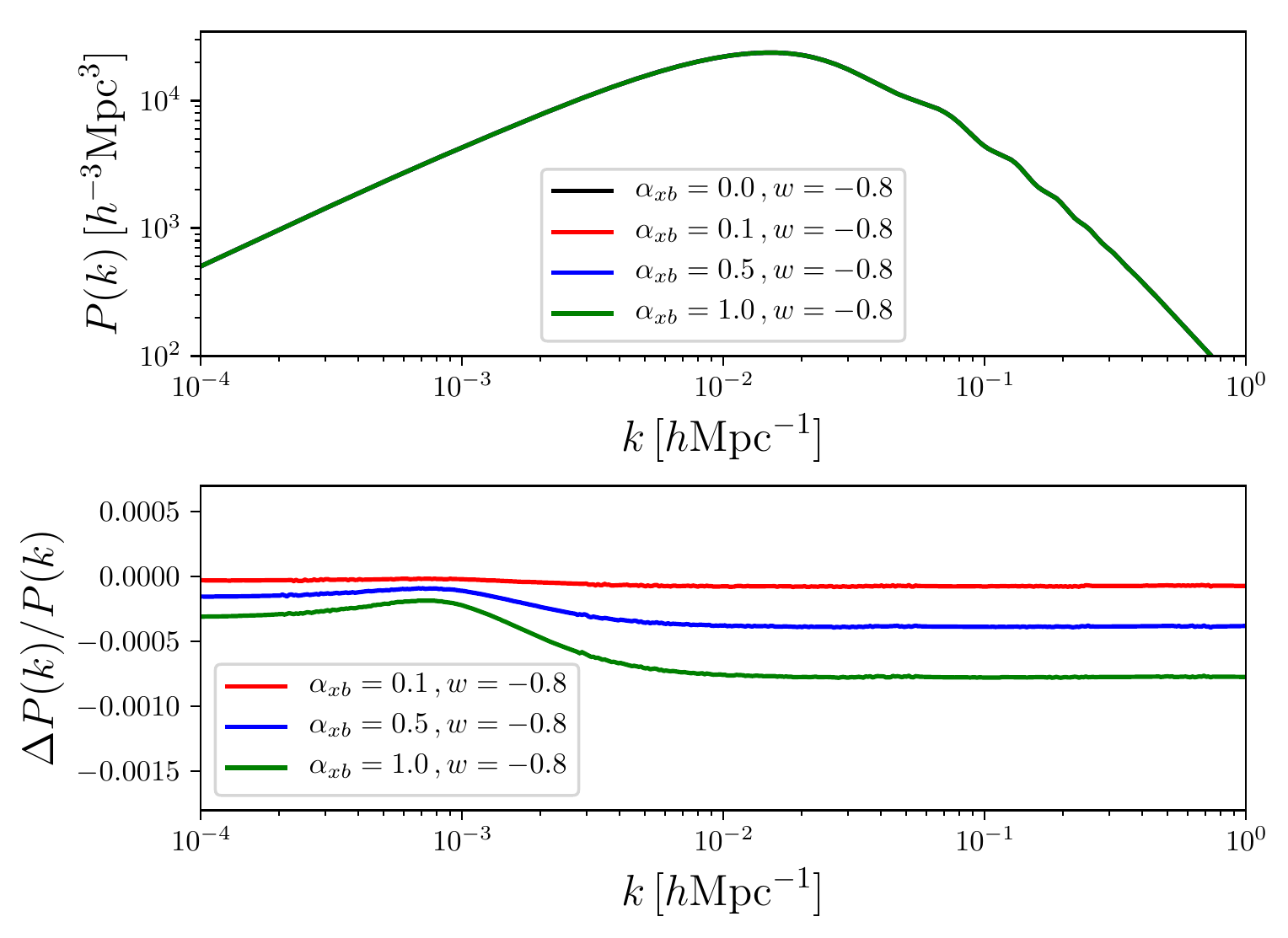}
\caption{As in Fig.~\ref{fig:fig1} but considering the matter power spectrum at redshift $z=0$. Note that we have plotted the linear power spectrum. Again the differences between the curves in the upper panel are too small to be appreciated by the naked eye. The decrease in power seen over all scales is due to a slowing down of structure growth, due to the DE-baryon drag term and leading to a minuscule decrease in $\sigma_8$, as we explain in Sec.~\ref{sec:physicalexplanation}.}
\label{fig:fig2}
\end{figure}

Aside from affecting the CMB temperature power spectrum, DE-baryon interactions are also expected to affect the clustering of the LSS, thus possibly leaving a signature in the power spectrum of matter fluctuations. We thus check the effect of increasing $\alpha_{xb}$ on the matter power spectrum $P(k)$ at redshift $z=0$. Our results are shown in Fig.~\ref{fig:fig2}. Again, as the upper panel of Fig.~\ref{fig:fig2} shows, the effect of $\alpha_{xb}$ on $P(k)$ leads to changes which are not distinguishable by the naked eye. In fact, in the lower panel of Fig.~\ref{fig:fig2} we show the relative changes of the power spectra within the interacting DE-baryon cosmologies with respect to the baseline case of $\alpha_{xb}=0$, and find these changes to be extremely tiny, with $\Delta P(k)/P(k)<0.1\%$ over most of the wavenumber range. These changes are too tiny to be observable by current or near-future LSS surveys. We expect these changes to be due to an overall suppression of structure growth due to the DE-baryon drag, which should effectively lead to a lower value of $\sigma_8$. We will discuss this further in Sec.~\ref{sec:physicalexplanation}.

Note that in Fig.~\ref{fig:fig2} the quantity plotted is the linear matter power spectrum. Therefore, the plots are truly reliable only up to $k_{\rm max} \approx 0.1\,h{\rm Mpc}^{-1}$, the approximate non-linear wavenumber today. Non-linear corrections to the power spectrum are typically computed using \texttt{Halofit} (which essentially consists of a fitting function calibrated to N-body simulations) or more generally emulating N-body simulations including hydrodynamic and baryonic effects. Due to the fact that no existing N-body simulation includes the effect of the DE-baryon scattering, which to the best of our knowledge is being studied here for the first time, we have no way of estimating how non-linear corrections behave within our model. In particular, it is possible that the contribution of baryonic effects to non-linear corrections might be strongly affected by the DE-baryon scattering: however, there is currently no way of telling. The only way to settle this issue is to run dedicated simulations, which is way beyond the scope of this work. However, should it turn out that structure formation in the non-linear regime is strongly affected by DE-baryon scattering, one could hope to probe such a scattering process using measurements of non-linear clustering in the large-scale structure, a possibility otherwise precluded by other cosmological probes as we are showing in this work. We hope to address this issue in a follow-up work.

\begin{figure}
\includegraphics[width=0.5\textwidth]{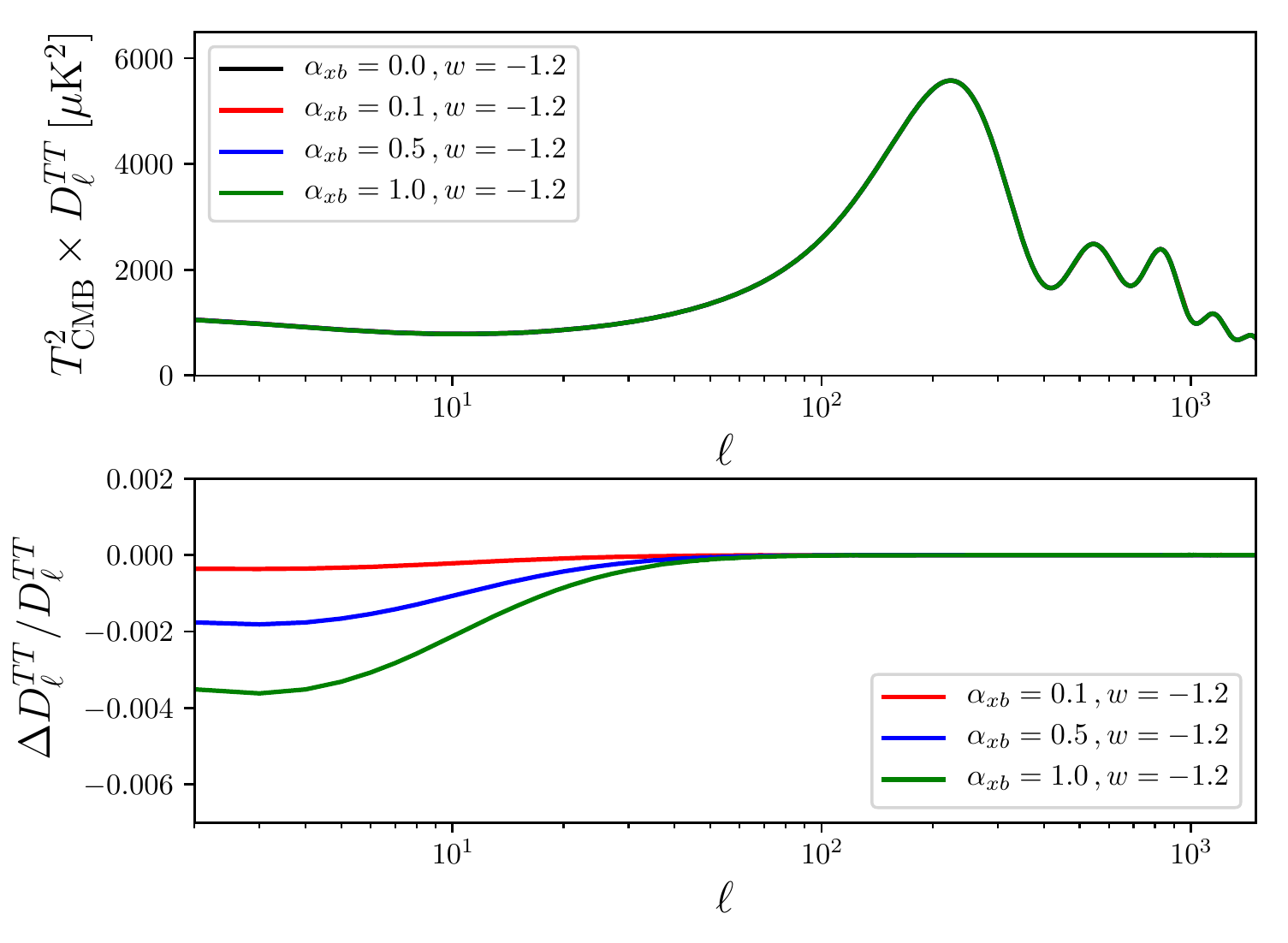}
\caption{As in Fig.~\ref{fig:fig1}, but considering a phantom DE component with EoS fixed to $w_x=-1.2$. Again the differences between the curves in the upper panel are too small to be appreciated by the naked eye. In this case the decrease in power at low multipoles is due to a suppressed late-time integrated Sachs-Wolfe effect, as we explain in Sec.~\ref{sec:physicalexplanation}.}
\label{fig:fig3}
\end{figure}

\begin{figure}
\includegraphics[width=0.5\textwidth]{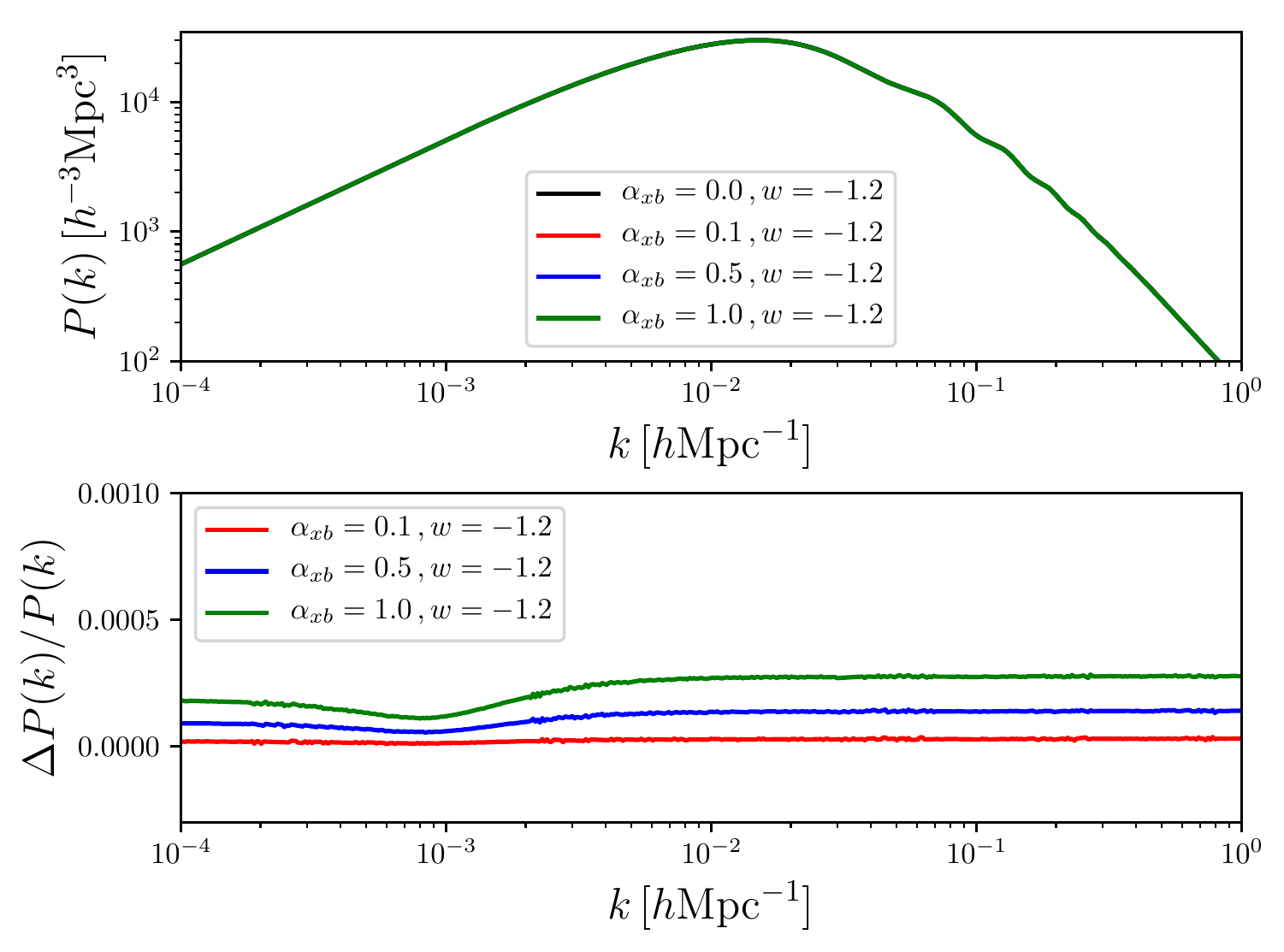}
\caption{As in Fig.~\ref{fig:fig2} but considering a phantom DE component with EoS fixed to $w_x=-1.2$. Again the differences between the curves in the upper panel are too small to be appreciated by the naked eye. The increase in power seen over all scales is due to a speeding up of structure growth, due to the DE-baryon drag term and leading to a minuscule increase in $\sigma_8$, as we explain in Sec.~\ref{sec:physicalexplanation}.}
\label{fig:fig4}
\end{figure}

We now move on to the case of a phantom DE component.~\footnote{We remark that phantom DE components are typically problematic from a theoretical point of view, due to their violating the strong energy condition, which leads to instabilities~\citep{Vikman:2004dc,Sawicki:2012pz}. Nonetheless, it is in principle possible to obtain phantom DE components which are effectively stable, for instance within modified gravity models~\citep[see e.g.][]{Elizalde:2004mq,Jhingan:2008ym,Setare:2008mb,Deffayet:2010qz,Cognola:2016gjy,Dutta:2017fjw,Casalino:2018tcd}.} As for the quintessence-like DE case, for purely pedagogical purposes we fix $w_x=-1.2$. Examining Eq.~(\ref{eq:thetab}), we can expect the effects of the DE-baryon scattering to be comparable in magnitude to those we found in the quintessence-like case, albeit reversed in sign. We confirm that this is indeed the case in Fig.~\ref{fig:fig3}, where we plot the resulting CMB temperature power spectra and relative deviations from the baseline case of $\alpha_{xb}=0$ for the same choices of $\alpha_{xb}$ as in Fig.~\ref{fig:fig1}, and in Fig.~\ref{fig:fig4} where we do the same for the matter power spectrum. We again expect these changes to be due to a reduced LISW effect and an overall enhancement of structure growth, and shall comment more on these effects in the following section.

\section{Physical explanation of results}
\label{sec:physicalexplanation}

We now turn to explain the results we found in the previous sections, summarized in Figs.~\ref{fig:fig1},\ref{fig:fig2},\ref{fig:fig3},\ref{fig:fig4}. There are two key questions we need to address. The first one is: what is the physics underlying the changes we found? The second one is: why, somewhat counter-intuitively, are the changes so small even for so large cross-sections? After addressing these questions, we study whether the DE-baryon scattering signatures are degenerate with signatures of the DE sound speed, which would complicate the prospects of identifying DE-baryon scattering signatures even further, finding that the answer is fortunately no.

Let us first consider the physics responsible for the changes in the CMB temperature anisotropy power spectrum (lower panels of Fig.~\ref{fig:fig1} and Fig.~\ref{fig:fig3}). Earlier, we raised the suspicion that these minuscule changes were due to changes in the LISW effect, with the direction of these changes being dependent on whether $w_x>-1$ (enhanced LISW effect) or $w_x<-1$ (suppressed LISW effect). Recall first of all that the integrated Sachs-Wolfe effect is a source of secondary anisotropies in the CMB, and is driven by time-variations in the gravitational potentials, which can only be present when the Universe is not matter-dominated~\citep{Sachs:1967er}. To linear order in temperature perturbations, the contribution of the LISW effect to temperature anisotropies $\Theta$ at a multipole $\ell$ from a mode with wavenumber $k$, $\Theta_{\ell}^{\rm LISW}$, is given by:
\begin{eqnarray}
\Theta_{\ell}^{\rm LISW}(k) = \int_{\eta_1}^{\eta_0}d\eta\,e^{-\tau(\eta)} \left [ \dot{\Psi}(k,\eta)-\dot{\Phi}(k,\eta) \right ]j_{\ell}(k(\eta_0-\eta))\,,
\label{eq:lisw}
\end{eqnarray}
where $\Psi$ and $\Phi$ are the two Newtonian gravitational potentials (note that in writing Eq.~(\ref{eq:lisw}) we have temporarily switched to the Newtonian gauge), $\tau$ is the optical depth, $\eta_0$ is the current conformal time, $\eta_1$ is the conformal time at an arbitrary point in time well into the matter-domination era, and $j_{\ell}$ is the Bessel function of order $\ell$. Therefore, LISW contributions to the CMB anisotropies are important only when DE starts dominating, making the gravitational potentials decay.

\begin{figure}
\includegraphics[width=0.5\textwidth]{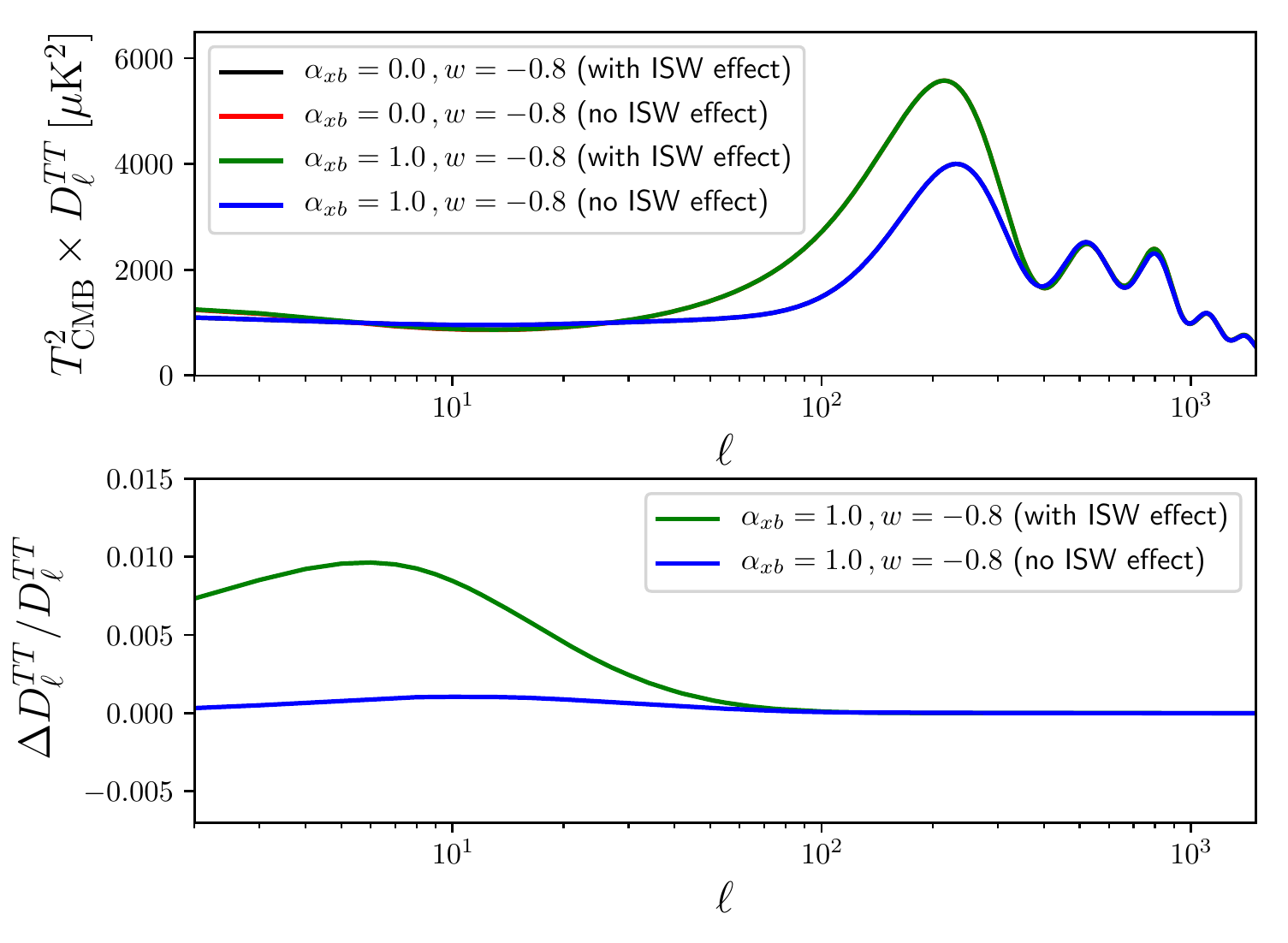}
\caption{Impact of increasing the Thomson ratio $\alpha_{xb}$ on the CMB temperature power spectrum, highlighting the contribution from the integrated Sachs-Wolfe (ISW) effect. \textit{Upper panel}: CMB temperature anisotropy power spectra for $\alpha_{xb}=0$ with (black curve) and without the ISW source term (red curve) and $\alpha_{xb}=1$ with (green curve) and without the ISW source term (blue curve), with the DE EoS fixed in all the cases to $w_x=-0.8$. \textit{Lower panel}: in the upper panel, the differences between the green curve and the black curve, as well as between the blue curve and the red curve, are too small to be seen by the naked eye. Therefore, we plot the relative change in power between the $\alpha_{xb}=1$ and baseline $\alpha_{xb}=0$ cosmologies, for the case where the ISW source term is included (green curve), and for the case where it is removed (blue curve). For clarity, we remark that the lower panel green curve shows the relative difference between the upper panel green and black curves, and similarly the lower panel blue curve shows the relative difference between the upper panel blue and red curves. The figure shows how the changes in the CMB temperature power spectrum coming from the DE-baryon scattering are entirely due to a variation in the ISW effect (since once the ISW effect is removed there are no residual changes in cosmologies with different $\alpha_{xb}$), and more precisely in the late ISW (LISW) effect, as discussed in Sec.~\ref{sec:physicalexplanation}.}
\label{fig:fig5}
\end{figure}

To confirm that the relative changes we are seeing in the lower panels of Fig.~\ref{fig:fig1} and Fig.~\ref{fig:fig3} are indeed due to a modified LISW effect, we consider again the comparison between a cosmology with $\alpha_{xb}=1$ and $\alpha_{xb}=0$ at fixed $w_x=-0.8$ (the former corresponding to the green curve in Fig.~\ref{fig:fig1}), but this time switching off the ISW source term in \texttt{CAMB}. In the upper panel of Fig.~\ref{fig:fig5}, we plot the CMB power spectra for the four resulting cosmologies: $\alpha_{xb}=1$ with (green curve) and without (blue curve) ISW source term, and similarly $\alpha_{xb}=0$ with (black curve) and without (red curve) ISW source term. In the lower panel of Fig.~\ref{fig:fig5} we then plot the relative changes between the $\alpha_{xb}=1$ and $\alpha_{xb}=0$ power spectra for the case with (green curve) and without (blue curve) ISW source term (note therefore that the green curve in Fig.~\ref{fig:fig5} corresponds to the green curve in Fig.~\ref{fig:fig1}). As is clear from Fig.~\ref{fig:fig5}, once the ISW source term is removed, the residual changes between cosmologies with different $\alpha_{xb}$ are virtually erased.~\footnote{Note that the tiny remaining changes barely visible by eye in the blue curve in the lower panel of Fig.~\ref{fig:fig5} are well below the precision of \texttt{CAMB} and hence are completely compatible with numerical noise.} A completely analogous plot, not shown here for the sake of conciseness, is obtained when considering a phantom DE component. The results shown in Fig.~\ref{fig:fig5} perfectly agree with our interpretation of the changes induced in the CMB temperature anisotropy power spectrum by increasing the Thomson ratio $\alpha_{xb}$ as being completely due to a change in the late-time integrated Sachs-Wolfe effect, and hence to changes in the behaviour of the decaying gravitational potentials during the dark energy-domination era.

Let us focus on the quintessence-like case, $w_x>-1$. As we have seen in Fig.~\ref{fig:fig1} and Fig.~\ref{fig:fig5}, the presence of a DE-baryon interaction has led to an enhanced LISW effect. From Eq.~(\ref{eq:lisw}) we see that an increased time variation in the Newtonian potentials (more precisely, an increased potential decay, given that potentials decay during the dark energy-dominated era) leads to an enhanced LISW effect. We expect this to be due to a decrease in the DE perturbations. In fact, it is well known~\citep[see e.g.][]{Weller:2003hw,Calabrese:2010uf} that DE perturbations help preserving potentials (in other words, they obstruct the potential decay). Hence, reducing the DE perturbations eases the potential decay and enhances the LISW effect. From a mathematical point of view, the DE-baryon scattering term in Eq.~(\ref{eq:thetade}) tends to increase the DE velocity, $\theta_x$. In the quintessence-like DE case where $w_x>-1$, the sign of the term proportional to $\theta_x/k^2$ on the right-hand side of the DE density contrast equation, Eq.~(\ref{eq:deltade}), is such that this increase in $\theta_x$ leads to a decrease in $\delta_x$ with respect to the baseline case of no interactions. The decrease of $\delta_x$ during the DE domination epoch leads to an increased decay of the gravitational potentials, and hence to an enhanced LISW effect.

In the phantom-like case where $w_x<-1$, similar considerations hold, albeit with the net effect on $\delta_x$ being reversed due to the presence of the $(1+w_x)$ factor in front of the $\theta_x/k^2$ term on the right-hand side of the DE density contrast equation, Eq.~(\ref{eq:deltade}). In this case, $\delta_x$ increases with respect to the baseline case of no interactions, helping in preserving the gravitational potentials from decay and thus reducing the LISW effect. There is another interesting point to be noted by comparing Fig.~\ref{fig:fig1} for the quintessence-like case with Fig.~\ref{fig:fig3} for the phantom-like case. We see that at a fixed deviation of the DE EoS from $w_x=-1$ ($\vert \Delta w_x \vert =0.2$), and at a fixed value of $\alpha_{xb}$, the magnitude of the DE-baryon scattering effect is larger for $w_x>-1$ than for $w_x<-1$. This finding is completely consistent with earlier findings in~\citep{Weller:2003hw}, who investigated the effect of DE perturbations on the ISW effect for both quintessence- and phantom-like DE. It was found that the magnitude of the ISW effect is larger for $w_x>-1$ than for $w_x<-1$, due to the different behaviour of DE perturbations (and in particular whether they are of the same sign of the matter perturbations) and of the decay of potentials in response to the different background behaviour of the DE component, whose energy density is increasing with the expansion for the phantom-like case, and decreasing for the quintessence-like case. We refer the reader to~\citep{Weller:2003hw} for more details.

Having explained the physics underlying the changes in the CMB temperature anisotropy power spectrum due to the scattering between DE and baryons, we now turn to examine the shifts induced by these processes on the matter power spectrum. We saw earlier that, in the quintessence-like case $w_x>-1$, the net effect of the DE-baryon scattering was to suppress DE perturbations, hence easing the decay of gravitational potentials and enhancing the LISW effect. In the case of the matter power spectrum, we expect that the increased decay of the gravitational potentials is going to slightly suppress structure formation, as baryon and cold DM overdensities are related to the gravitational potentials via the Poisson equation. In addition, DE-baryon scattering leads to a drag term in the baryon velocity equation, Eq.~(\ref{eq:thetab}), which also slows down the growth of structure. This is analogous to the DE-DM drag studied in~\citep{Simpson:2010vh} in the context of DE-DM scattering. However, the effect we find is much smaller than that found by~\citep{Simpson:2010vh} simply because baryons are subdominant compared to the DM, with the latter dominating the structure formation dynamics. We therefore expect our effects on the matter power spectrum to be suppressed approximately by a factor of $(\Omega_b/\Omega_c)^2$, which is a quantity of ${\cal O}(10^{-2})$, with respect to the effects found by~\citep{Simpson:2010vh}.

The combination of the two effects described earlier (the increased decay of the gravitational potentials also responsible for the enhanced LISW effect, and the drag term in the baryon velocity equation) lead to an overall suppression of the matter power spectrum, which we observe in Fig.~\ref{fig:fig2}. In addition, we expect the effects of the DE-baryon scattering to lead to a slightly lower value of $\sigma_8$. Indeed, we find this to be the case, although the induced changes are tiny. For $\alpha_{xb}=1$, we find that the differences in $\sigma_8$ with respect to the baseline case with $\alpha_{xb}=0$ occur at the fourth decimal digit, and hence are observationally impossible to detect even with the most ambitious future LSS surveys~\citep[see e.g.][]{Amendola:2012ys,2012arXiv1211.0310L,2015arXiv150303757S,2016arXiv161100036D}. Obviously, such a small reduction of $\sigma_8$ is also completely unable to address the tension between CMB and low-redshift (redshift-space distortions and cosmic shear) determinations of $\sigma_8$~\citep{DiValentino:2018gcu}. In the phantom-like case $w_x<-1$, we expect the effects of DE-baryon scattering on the matter power spectrum to be reversed in sign, due to the reduced decay of the gravitational potentials, as well as the drag term in the baryon velocity equation appearing with the opposite sign. These two effects act to enhance structure growth, and hence the amplitude of the matter power spectrum. This explains the results we illustrate in Fig.~\ref{fig:fig4}.

In conclusion, we find that for $\alpha_{xb} \lesssim 1$, the reduction or enhancement of power in the matter power spectrum (depending on whether DE is quintessence-like or phantom-like) is too small to ever be detected. Moreover, such a suppression is expected to be strongly degenerate with both $\sigma_8$ and the (possibly scale-dependent) bias of the LSS tracer in question (recall that we have considered the matter power spectrum, but in reality one observes the power spectrum of LSS tracers, which is biased with respect to the underlying matter power spectrum), although the latter degeneracy might be mitigated by jointly considering cross-correlations between LSS clustering and CMB lensing~\citep[see e.g.][]{Giusarma:2018jei}. Moreover, we also expect the effects of the DE-baryon scattering to be completely negligible both in polarization and CMB lensing. We have checked this explicitly by computing the E-mode, B-mode, and CMB lensing power spectra for the cosmologies discussed in this section, and found them to be even smaller than those we have observed in the temperature and matter power spectra, and hence virtually undetectable. For the sake of conciseness, we do not show these results here.

Having addressed the question of what is the physics responsible for the changes we found in the CMB and matter power spectra, we now turn to the question of why such changes are so tiny even when considering a huge DE-baryon cross-section, $\alpha_{xb} \sim {\cal O}(1)$, comparable to nuclear cross-sections. In order to leave a significant imprint on the CMB temperature anisotropy power spectrum, DE-baryon scattering should impact the pre-recombination dynamics of the baryon fluid, and hence generate significant primary anisotropies. However, it is clear that from Eq.~(\ref{eq:thetab}) that, even considering $\alpha_{xb} \sim {\cal O}(1)$ and hence $\sigma_{xb} \simeq \sigma_T$, the effect of DE-baryon scattering is completely subdominant with respect to that of Thomson scattering as it is suppressed by two terms. The first is the ratio $\rho_x/\rho_b \ll 1$, reflecting the fact that DE is completely subdominant at the pre-recombination epoch, and hence there is no target for DE-baryon scattering to occur. The second is the $(1+w_x)$ term, which suppresses the effect of DE-baryon scattering the more DE behaves as a cosmological constant, as indicated by data.

Therefore, for $\alpha_{xb} \ll 1$, the only possibility for DE-baryon scattering to leave an imprint on the CMB is through secondary anisotropies, such as the LISW effect as we have seen earlier. However, these effects are again strongly suppressed by the fact that $w_x$ is close to $-1$ (the closer $w_x$ is to $-1$, the smaller DE perturbations are) and that baryons are strongly subdominant at late times, when it is DM that is playing the dominant role in structure formation. As argued earlier, this also explains why the effects of DE-baryon scattering on the matter power spectrum are so small.

These considerations hold even at late times, when the ratio $\rho_x/\rho_b$ is no longer small, but of ${\cal O}(10)$. Note that this ratio only enters the baryon velocity equation, Eq.~(\ref{eq:thetab}), but not the DE velocity equation, Eq.~(\ref{eq:thetade}). The effect of DE-baryon scattering on the CMB is still small since the process can only generate secondary anisotropies. More importantly, these secondary anisotropies (such as the LISW effect) are governed by the behaviour of gravitational potentials: their evolution is mostly controlled by the DE overdensity and velocity, whose governing equations do not contain the $\rho_x/\rho_b$ factor.

With regards to the matter power spectrum, it is true that the term $\rho_x/\rho_b$ could in principle affect the baryon velocity equation in a significant way (note, however, that the matter power spectrum is directly sensitive to the baryon overdensity, whose main source term at late times comes from the metric perturbations $h$). However, baryons contribute a subdominant fraction to gravitational potentials. To put it differently, the contribution of baryons to the linear matter transfer function is suppressed by $\Omega_b/\Omega_c$, with $\Omega_b$ and $\Omega_c$ the baryon and cold dark matter density parameters. Therefore, the contribution of baryons to the linear matter power spectrum is suppressed by a factor $(\Omega_b/\Omega_c)^2$. The suppression factor is of ${\cal O}(0.01)$, so it acts to offset the potential large effect of the $\rho_x/\rho_b$ factor on the matter power spectrum. This is true at linear level, but we could expect that at the non-linear level DE-baryon scattering could significantly modify the baryon transfer function, possibly leading to visible effects. This further motivates a follow-up work devoted to running dedicated N-body simulations to study the effect of DE-baryon scattering in the non-linear regime.

\subsection{A possible degeneracy with the dark energy sound speed}
\label{subsec:soundspeed}

We have already seen that the effect of DE-baryons scattering on cosmological observables is tiny for reasonable values of the DE-baryon cross-section and this alone would be enough to conclude that these effects are virtually undetectable not only with current surveys but also with future ones. One may also worry that the effects of DE-baryon scattering might be degenerate with those of other cosmological parameters, such as $\sigma_8$ in the case of the matter power spectrum. In the case of the CMB power spectrum, the fact that DE-baryon scattering only affects the LISW effect raises the suspicion that $\alpha_{xb}$ might be degenerate with some other DE property which leaves comparable imprints on the LISW effect, such as the DE sound speed squared, $c_{s,x}^2$, which earlier we fixed to $c_{s,x}^2=1$. In fact, from the discussions in~\citep{Calabrese:2010uf} and especially Fig.~1 therein, one might legitimately suspect that the imprint of $\alpha_{xb}$ on the LISW effect would be completely degenerate with that of $c_{s,x}^2$.

Physical values of the DE sound speed should lie in the region $0 \leq c_{s,x}^2 \leq 1$. Outside of this region one faces tachyonic/gradient instabilities and/or instabilities connected to superluminal propagation. In the standard scenario, one fixes $c_{s,x}^2=1$. Therefore, we need to check what is the impact of reducing the DE sound speed squared to values as low as $0$, and whether the resulting effects on the CMB can be mimicked by DE-baryon scattering. One important point to note is that in order to study the effect of the DE sound speed, one must again consider values of the DE EoS $w_x \neq -1$, since a cosmological constant has no perturbations and hence a cosmology with DE in the form of a cosmological constant has no sensitivity to the DE sound speed. The impact of the DE sound speed on the CMB power spectrum has been discussed in detail in~\citep{Calabrese:2010uf}. There it was found that the effect of decreasing $c_{s,x}^2$ from $1$ to $0$ is to suppress the LISW effect when the DE EoS satisfies $w_x>-1$, but for $w_x<-1$ decreasing the sound speed results in an enhancement of the LISW effect. Heuristically, at least for the case where $w_x>-1$, this can be understood as follows: the more we decrease $c_{s,x}^2$, the more DE can cluster and effectively behave as ``cold'' dark energy. Clustering enhances the DE perturbations, which as discussed earlier protect the potentials from decaying, thus leading to a smaller contribution to the LISW effect as can be seen from Eq.~(\ref{eq:lisw}). We refer the reader to~\citep{Calabrese:2010uf} for a mathematically rigorous discussion of the effect of $c_{s,x}^2$ on the CMB. See also e.g.~\citep{DeDeo:2003te,Hannestad:2005ak,Xia:2007km,Mota:2007sz,dePutter:2010vy,Carbone:2010sb,Archidiacono:2014msa} for further works examining the effect of $c_{s,x}^2$ on cosmological observations.

\begin{figure}
\includegraphics[width=0.5\textwidth]{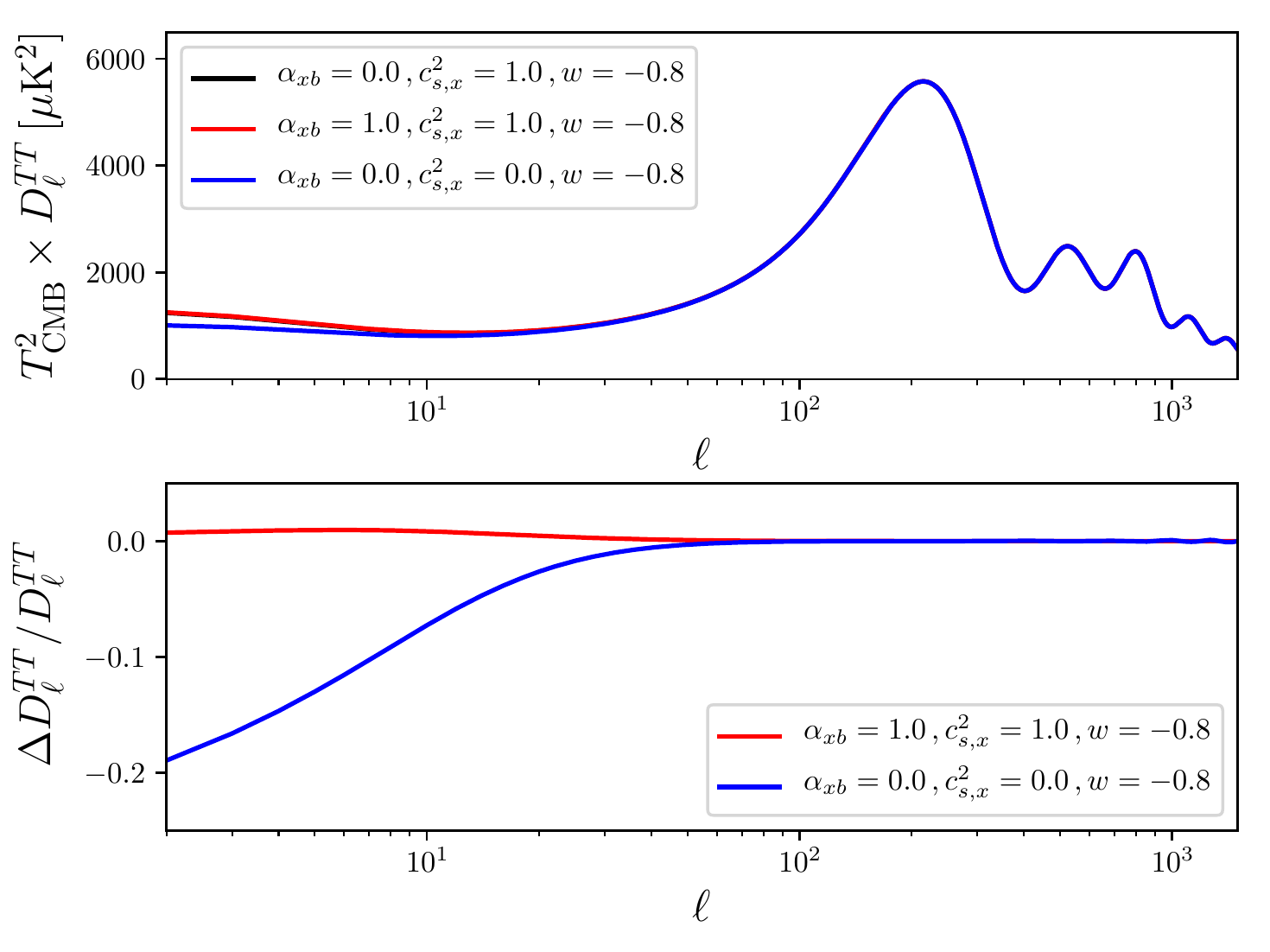}
\caption{Comparison of the effects of the DE-baryon scattering and of changing the DE sound speed squared $c_{s,x}^2$. \textit{Upper panel}: CMB temperature anisotropy power spectra for $\alpha_{xb}=0$ and $c_{s,x}^2=1$ (black curve), $\alpha_{xb}=1.0$ and $c_{s,x}^2=1.0$ (red curve), and $\alpha_{xb}=0$ and $c_{s,x}^2=0$ (blue curve). All cosmological parameters are fixed to their best-fit values given the \textit{Planck} 2018 results~\citep{2018arXiv180706209P}, with the exception of the DE EoS which is fixed to $w_x=-0.8$, in the quintessence region. \textit{Lower panel}: relative change in power with respect to the baseline model with $\alpha_{xb}=0.0$ and $c_{s,x}^2=1.0$, with the same color coding as above. In both cases the relative changes are due to an ehanced late-time integrated Sachs-Wolfe effect in the case where $\alpha_{xb}$ is increased, or to a suppression in the same effect when $c_{s,x}^2$ is decreased. Similar effects, although reversed in sign, are observed when considering a phantom DE component (not shown here).}
\label{fig:fig6}
\end{figure}

\begin{figure}
\includegraphics[width=0.5\textwidth]{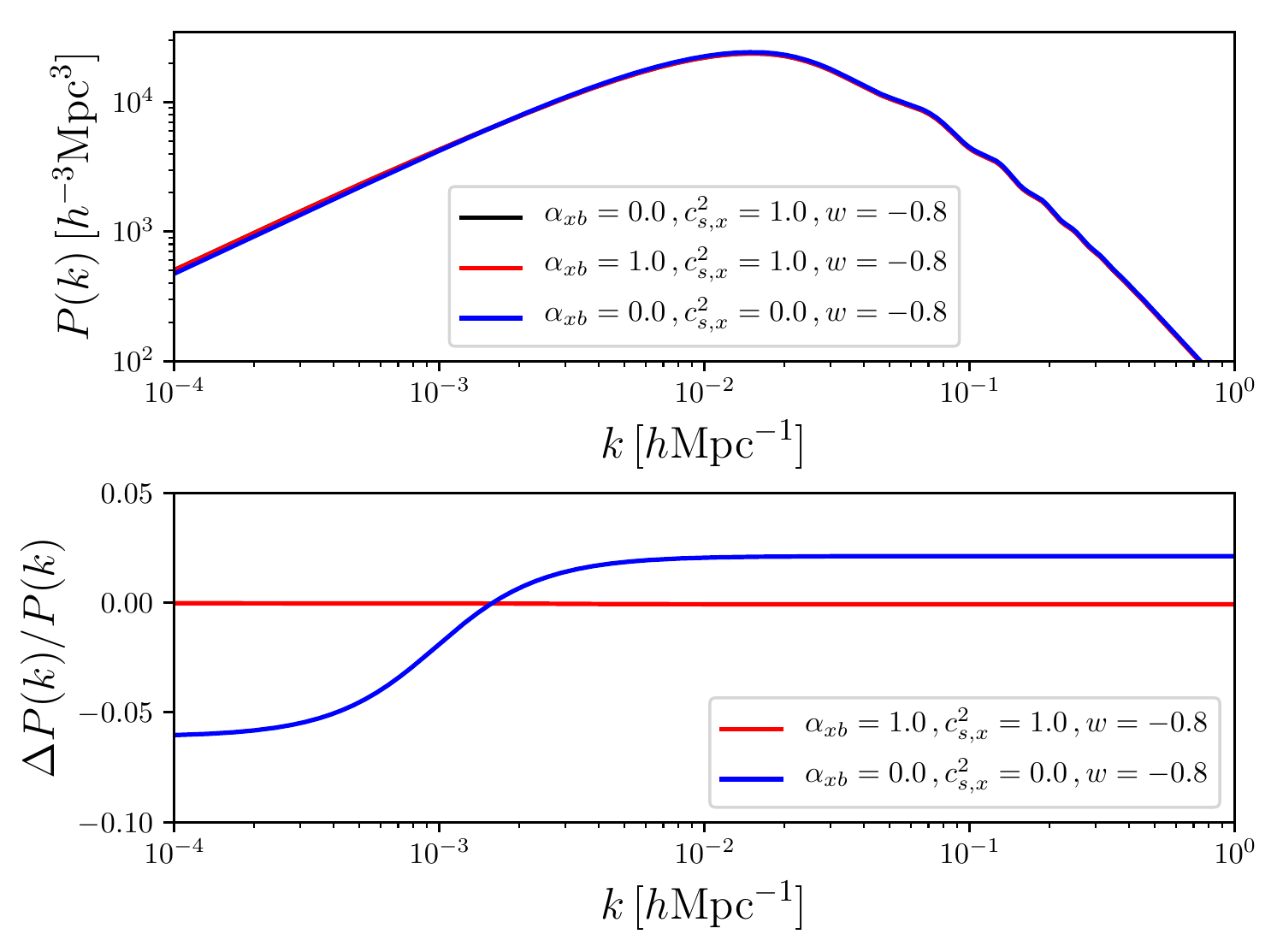}
\caption{As in Fig.~\ref{fig:fig6} but considering the linear matter power spectrum at redshift $z=0$. Similar effects, although reversed in sign, are observed when considering a phantom DE component (not shown here).}
\label{fig:fig7}
\end{figure}

To understand whether the effect of $c_{s,x}^2$ can be to some extent degenerate with that of $\alpha_{xb}$, we again consider a baseline cosmology with $\alpha_{xb}=0$, $c_{s,x}^2=1$, and $w_x=-0.8$. We then compare the resulting CMB and matter power spectra with the power spectra of the cosmology with $\alpha_{xb}=1$ and $c_{s,x}^2=1$ we already discussed earlier (Fig.~\ref{fig:fig1} and Fig.~\ref{fig:fig2}), as well as with those for a cosmology with $\alpha_{xb}=0$ and $c_{s,x}^2=0$. We show our results in Fig.~\ref{fig:fig6} for the CMB temperature anisotropy power spectrum, and in Fig.~\ref{fig:fig7} for the matter power spectrum. As is clear from the lower panels of these figures, where we plot the relative changes with respect to the baseline cosmology, the effects of DE-baryon scattering and the DE sound speed on both probes are actually quite distinct. In particular, in the case of the CMB as shown in Fig.~\ref{fig:fig6}, this easily follows from the discussions in~\citep{Calabrese:2010uf}, where decreasing $c_{s,x}^2$ for $w_x>-1$ was found to lead to a suppressed LISW effect as discussed above, whereas increasing $\alpha_{xb}$ for $w_x>-1$ leads to an enhanced LISW effect as we discussed in Sec.~\ref{sec:physicalexplanation}. The effects are opposite in sign and hence, at least in principle, distinguishable, barring the fact that both are extremely small and show up on very large scales where cosmic variance completely swamps the signal. The effect of the DE sound speed on the matter power spectrum as shown in Fig.~\ref{fig:fig7} is also quite distinct with respect to that of DE-baryon scattering, since the former changes sign at intermediate scales, whereas the latter gives a nearly scale-independent suppression. Similar conclusions hold when we consider a phantom-like DE component with $w_x<-1$, with all effects reversing sign but being, in principle, distinguishable between each other.

\section{Conclusions}
\label{sec:conclusions}

In this paper, we have considered the possibility that dark energy (DE) and baryons might scatter off each other. We have quantified the strength of the DE-baryon scattering by the dimensionless parameter $\alpha_{xb}$, given by the ratio of the DE-baryon cross-section to the Thomson cross-section. We have argued that, on purely general grounds, we expect $\alpha_{xb} \ll 1$ from non-cosmological probes, given the non-observation of DE in colliders or precision tests of gravity. This does not exclude the possibility that one might construct a specific UV complete model wherein $\alpha_{xb} \sim {\cal O}(1)$ or larger is allowed, appropriately screened, and consistent with all experimental tests, albeit highly challenging. The question we have addressed in this paper is then: what signatures would a DE-baryon scattering leave on cosmological observables?

We have found, surprisingly, that even for $\alpha_{xb} \sim {\cal O}(1)$ or larger, DE-baryon scattering leaves minuscule imprints on the CMB temperature anisotropy power spectrum and the matter power spectrum. The size of these imprints also depends on how much the DE equation of state $w_x$ deviates from that of a cosmological constant, with the direction of these imprints depending crucially on whether DE is quintessence-like ($w_x>-1$) or phantom-like ($w_x<-1$). Considering a quintessence-like DE component, we have found that the effect of DE-baryon scattering is to decrease the DE perturbations, which in turn eases the decay of gravitational potentials at late times. This leads to an enhanced late-time integrated Sachs-Wolfe effect, and suppresses the late-time growth of structure. These effects show up as an enhancement of power in the CMB low-$\ell$ tail, as well as a nearly scale-independent suppression of the matter power spectrum $P(k)$, leading to a tiny reduction in $\sigma_8$. For a phantom-like DE component, the sign of all these effects are reversed, thus leading to a suppressed LISW effect and an enhanced matter power spectrum.

We have found that for $\alpha_{xb} \sim {\cal O}(1)$, all these effects are too small to be observable both in current and future surveys. For the CMB, DE-baryon scattering leads to sub-$\%$ changes at very low $\ell$, in a regime where cosmic variance completely hinders the possibility of detecting such a signal. In the matter power spectrum the signatures are more than an order of magnitude smaller than those in the CMB, and hence well below the projected uncertainty of even the most optimistic future LSS survey. We have also studied whether the signatures of DE-baryon scattering might be degenerate with signatures of the DE sound speed: we have showed that the two effects are quite distinct both on the CMB (where they work in opposite directions) and on the matter power spectrum, and hence in principle distinguishable if it were not for the fact that both are extremely tiny.

In conclusion, we have confirmed the suspicion raised in~\citep{Simpson:2010vh} that huge interaction cross-sections between dark energy and baryons are allowed without disrupting the CMB or structure formation. We remind the reader that $\alpha_{xb} \sim {\cal O}(1)$ corresponds to barn-scale cross-sections, which are extremely large and comparable to nuclear cross-sections. For comparison, current limits on dark matter-baryon scattering from DM direct detection experiments can be as constraining as $10^{-22}\,{\rm b}$ (\textit{i.e.} $10^{-46}\,{\rm cm}^2$) depending on the DM mass. Therefore, while alluring, the prospect of cosmological direct detection of dark energy appears to be a remote one, unless Nature has endowed DE and baryons with a huge interaction cross-section well above the barn scale, and managed to make it surpass all non-cosmological searches. One important caveat is that our analysis was performed at linear order in perturbation theory. It is possible that baryonic effects on non-linear corrections to the matter power spectrum might carry a visible imprint of DE-baryon scattering. The only way to find out for sure is to run dedicated N-body simulations. This is beyond the scope of the current paper, and we hope to address this issue in future work. Should non-linear effects significantly enhance the imprint of DE-baryon scattering on the matter power spectrum, it might be possible to probe such scattering by studying the clustering of the large-scale structure in the non-linear regime.

There are in principle avenues for further exploration along this line. The most strongly motivated follow-up work would look at running dedicated N-body simulations to study the effect of DE-baryon scattering in the non-linear regime and check whether the impact of baryonic effects on non-linear corrections to the matter power spectrum is significantly affected by DE-baryon scattering. On the model-building side, it would be interesting to try and construct a model or class of models featuring huge DE-baryon interactions on cosmological scales, which might thus leave visible imprints in cosmological observables, while still being consistent with terrestrial experiments. On the data analysis side, it could be interesting to investigate a DE-baryon interaction scenario with more freedom in the DE sector, for instance considering time-varying DE. In fact, it would be quite unnatural or at least surprising if a DE component featuring interactions with baryons had an equation of state constant in time. Furthermore, given how DE-baryon scattering leaves its largest signature in the CMB on large scales by affecting the LISW effect, it might be worth thinking about more efficient ways of isolating this signal, for instance by considering cross-correlations between temperature fluctuations from future CMB surveys~\citep[e.g.][]{2016arXiv161002743A,Ade:2018sbj,2019arXiv190708284T} and maps of overdensities in future LSS surveys~\citep[see e.g.][]{Amendola:2012ys,2012arXiv1211.0310L,2015arXiv150303757S,2016arXiv161100036D}, see for instance~\citep{Bean:2003fb,Hu:2004yd,Corasaniti:2005pq,Ho:2008bz,Ferraro:2014msa} for earlier works along this line of research. Likewise, if DE interacts with baryons but not with DM, this will lead to a ``baryon bias'' which might be constrained using motions of tidally disrupted stellar streams of merging galaxy clusters~\citep[see e.g.][]{Amendola:2001rc,Kesden:2006zb,Randall:2007ph}. Finally, if cosmic variance is the true killer of any prospect for cosmological direct detection of dark energy, it would be worth investigating whether it is possible to find signatures of DE-baryon scattering which can beat cosmic variance. Even if, as argued in our work, DE-baryon scattering affects structure formation ever so slightly, this will leave an imprint in the bias of LSS tracers such as galaxies. Since the bias of a given LSS tracer is not a random field, one can use it to beat cosmic variance provided a dense enough LSS tracer sample can be found~\citep[see e.g.][]{Seljak:2008xr,McDonald:2008sh,LoVerde:2016ahu}. If a DE-baryon scattering signal is truly imprinted in the LSS bias, it would be worth investigating whether one might ever be able to extract it. We shall investigate these and related issues in future work.

\section*{Acknowledgements}

We thank Philippe Brax and Fergus Simpson for useful discussions. S.V. is supported by the Isaac Newton Trust and the Kavli Foundation through a Newton-Kavli fellowship, and acknowledges a College Research Associateship at Homerton College, University of Cambridge. L.V. is supported through the research program ``The Hidden Universe of Weakly Interacting Particles'' with project number 680.92.18.03 (NWO Vrije Programma), which is partly financed by the Dutch Research Council. O.M. is supported by the Spanish grants FPA2017-85985-P and SEV-2014-0398 of MINECO, by PROMETEO/2019/083, and the European Union's Horizon 2020 research and innovation program under the Marie Sk\l odowska-Curie grant agreements 690575 and 674896. D.F.M. acknowledges support from the Research Council of Norway.

\bsp
\label{lastpage}
\end{document}